\documentclass[twocolumn,aps,showpacs,preprintnumbers,amsmath,amssymb,floatfix]{revtex4}

\usepackage{graphicx}
\usepackage{dcolumn}
\usepackage{bm}
\usepackage{color}

\begin{document}

\preprint{Hugger et al.}

\title{Magnetic barrier induced conductance fluctuations in quantum wires}

\author{S. Hugger} \altaffiliation[Present address: ]{Fraunhofer Institut f\"ur Angewandte
Festk\"orperphysik, Tullastr. 72, 79108 Freiburg, Germany.}\author{Hengyi Xu}\author{A. Tarasov}\author{M. Cerchez}\author{T. Heinzel}
\email{thomas.heinzel@uni-duesseldorf.de}
\affiliation{Heinrich-Heine-Universit\"at, Universit\"atsstr. 1,
40225 D\"usseldorf, Germany}
\author{I. V. Zozoulenko}
\affiliation{Department of Science and Technology, Link\"{o}ping
University, 60174 Norrk\"{o}ping, Sweden}
\author{D. Reuter}\author{A. D. Wieck}
\affiliation{Lehrstuhl f\"ur Angewandte Festk\"orperphysik, Ruhr-Universit\"at Bochum, 44780 Bochum, Germany
}
\date{\today}

\begin{abstract}
Quasi-ballistic semiconductor quantum wires are exposed to
localized perpendicular magnetic fields, also known as magnetic
barriers. Pronounced,
reproducible conductance fluctuations as a function of the magnetic barrier
amplitude are observed. The fluctuations are strongly temperature dependent and
remain visible up to temperatures of  $\approx
10\,\rm{K}$. Simulations based on recursive Green's functions suggest that the conductance
fluctuations originate from parametric interferences of the electronic wave functions which experience scattering
between the magnetic barrier and the electrostatic potential landscape.
\end{abstract}

\pacs{73.23.-b,75.70.Cn}
\maketitle

\section{\label{sec:1}INTRODUCTION}

Two-dimensional electron gases (2DEGs) exposed to inhomogeneous perpendicular magnetic fields show a wide variety of fascinating transport properties. \cite{Ye1995,Novoselov2002,Nogaret2003,Hara2004a} An elementary magnetic nanostructure is the \emph{magnetic barrier} (MB), i.e., a perpendicular magnetic field configuration which is strongly localized in the transport direction and homogeneous in the transverse direction. Theoretical studies \cite{Peeters1993,Avishai1990} have preceded experimental investigations of this system, which can be generated by placing the edge of a ferromagnetic film across a Hall bar containing the 2DEG and magnetizing the film along the transport direction. \cite{Monzon1997,Johnson1997} During the past ten years, a substantial quantity of theoretical studies has been published addressing various aspects of the magnetotransport properties of MBs. \cite{Ibrahim1997,Heide1999,Governale2000,Guo2002,Zhai2002,Zhai2005,Majumdar1996,Papp2001a,Papp2001b,Xu2001,
Lu2002,Jiang2002,Seo2004,Jalil2005,Zhai2006,Xu2007a} Magnetic barriers in quantum wires have been suggested as tunable spin filters, \cite{Majumdar1996,Papp2001a,Papp2001b,Xu2001,
Lu2002,Jiang2002,Seo2004,Jalil2005,Zhai2006} and it has been predicted that the conductance of such systems shows Fano resonances. \cite{Xu2007a} Furthermore, MBs should be capable of confining electrons in graphene sheets. \cite{Martino2007}

Despite this large body of theory, there have been relatively few experiments on MBs. \cite{Leadbeater1995,Monzon1997,Johnson1997,Kubrak2000,Hugger2007,Bae2007,Hong2007,Vancura2000,Cerchez2007} Up to now, all of them have been carried out in electron gases of width  $\geq 1\,\mathrm{\mu m}$ and could be explained within the semiclassical picture, whereas the majority of the theoretical results comprise quantum effects on MBs defined in quantum wires (QWRs). \cite{Majumdar1996,Papp2001a,Papp2001b,Xu2001,
Lu2002,Jiang2002,Seo2004,Jalil2005,Zhai2006,Xu2007a} Moreover, the well-known phenomenology of QWRs in homogeneous magnetic fields \cite{Beenakker1991} will be modified in such systems. For example, both the magnetoresistance peak due to boundary scattering \cite{Thornton1989} as well as the flux cancelation effect \cite{Beenakker1988a} should be suppressed, since they originate from electronic motion in spatially extended and homogeneous perpendicular magnetic field. Due to this state of the field, it is of great interest to perform transport experiments on MBs in preferably non-diffusive quantum wires.

Here, we report an investigation of the transport properties of quasi-ballistic quantum wires exposed to a magnetic barrier. Resistance fluctuations with a strongly temperature dependent amplitude are measured as a function of the barrier strength. These observations are interpreted within a recursive Green's function model as a manifestation of magnetic barrier - induced changes of the electronic interference pattern in the wire.

The outline of the paper is as follows. The sample preparation and the experimental setup are described in Sec. II. In Sec. III, the experimental results are reported and interpreted. The paper concludes with a summary in Sec. IV.

\section{\label{sec:2}SAMPLE PREPARATION AND EXPERIMENTAL SETUP}

\begin{figure}
\includegraphics{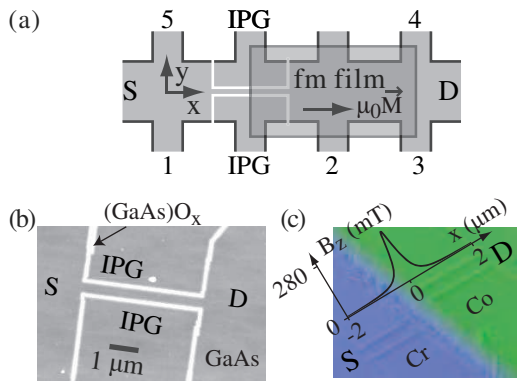}
\caption{(Color online) (a) Scheme of the sample layout. The QWR is formed by two oxide lines (white) in the Hall bar (light gray) and can be tuned by voltages applied to the in-plane gates (IPGs). The source and drain contacts are labeled by S and D, respectively.
The ferromagnetic film (dark gray) is magnetized in $x$-direction (magnetization $\mu_0\vec{M}$), and the magnetic barrier forms underneath the edges running in $y$-direction at the center of the QWR
and in the Hall cross, respectively. Scanning force microscope image of the sample with the QWR after the scanning probe lithography step (b) and of a dummy sample after deposition of the Cr layer and the ferromagnetic film (c), respectively. The overlay in (c) sketches the perpendicular magnetic field along the wire at a magnetization of the Co film of $\mu_0 M=1.1\,\mathrm{T}$.}
\label{MBCFinQWR_fig1}
\end{figure}

A $\mathrm{GaAs/Al_xGa_{1-x}As}$ heterostructure with a  2DEG residing  $55\,\mathrm{nm}$ below the surface was used for the experiments. The 2DEG has an electron density of $n=3.1\times 10^{15}\,\mathrm{m^{-2}}$ and a mobility of  $\mu=60\,\mathrm{m^2V^{-1}s^{-1}}$ at a temperature of $2.1\,\rm{K}$. The lateral layout of the samples is depicted in Fig. \ref{MBCFinQWR_fig1}. A Hall bar with Ohmic contacts has been prepared by conventional optical lithography.  Various QWR geometries  have been defined in the 2DEG by local oxidation with an atomic force microscope. \cite{Held1999}  Their lithographic width varies from $400\,\mathrm{nm}$ to $600\,\mathrm{nm}$, and their lengths from $1\,\mathrm{\mu m}$ to $9\,\mathrm{\mu m}$, respectively. The Fermi energy in the QWR can be tuned by voltages applied to the two in-plane gates (IPG). Subsequently, the structure was covered by a  Cr layer of $10\,\mathrm{nm}$  thickness, and one edge of a ferromagnetic film (Co or Dy,  thickness  $t=250\,\mathrm{nm}$) was aligned along the $y$-direction (i.e., perpendicular to the quantum wire) by electron beam lithography
and metallization at a base pressure of  $8\times 10^{-7}\,\mathrm{mbar}$. The opposite edge is located  at the center of a Hall cross, which allows measuring the film magnetization via Hall magnetometry.

The measurements were
performed in a $\mathrm{^4He}$ gas flow cryostat with a base temperature of $2\,\mathrm{K}$. The system is equipped with a
superconductive magnet that generates a homogeneous magnetic field
$B_h$, tunable between $-8\,\rm{T}$ and $8\,\rm{T}$.
The samples were mounted on a rotatable stage such that the orientation of $B_h$ could be adjusted between parallel to the
QWR ($x$-direction in Fig. \ref{MBCFinQWR_fig1})
and perpendicular to the 2DEG ($z$-direction). Parallel orientation with an accuracy of  $\pm 0.05^{\circ}$ is established by
measuring a Hall voltage of zero between contacts 1 and 5 for $B_{h}=8\,\rm{T}$.

\section{\label{sec:3}EXPERIMENTAL RESULTS AND INTERPRETATION}

Three samples of the geometry described above have
been measured, all showing a similar phenomenology.  Here, we present data from a $4\,\mathrm{\mu m}$ long
QWR with a Co film on top, acquired in three cooldowns.  Magnetotransport measurements as a function of  $B_{h}^{z}$ reveal that seven modes are occupied in this QWR, and we estimate its electronic width to $\approx 200\,\mathrm{nm}$.  As the Co film is magnetized in $x$-direction, the perpendicular component $B_{z}(x)$ of the fringe field forms the MB. Its shape  is given by \cite{Vancura2000}
\begin{equation}
B_{z}(B_{h}^{x},x)=-\frac{\mu_0 M(B_{h}^{x})}{4\pi}\ln
\frac{x^2+z^2_0}{x^2+(z_0+t)^2} \label{eqn1}
\end{equation}
where $\mu_0M$ denotes the magnetization of the Co film, and $z_0 =65\,\mathrm{nm}$ its distance to the 2DEG. At our maximum magnetization of  $\mu_0 M=1.1\,\mathrm{T}$,  Eq. \eqref{eqn1} gives a MB with a peak of $B_z(1\,\mathrm{T},x=0)\equiv B_{peak}=275\,\mathrm{mT}$ and a full width at half maximum of $290\,\mathrm{nm}$, as visualized in Fig. \ref{MBCFinQWR_fig1} (c).

\begin{figure}
\includegraphics{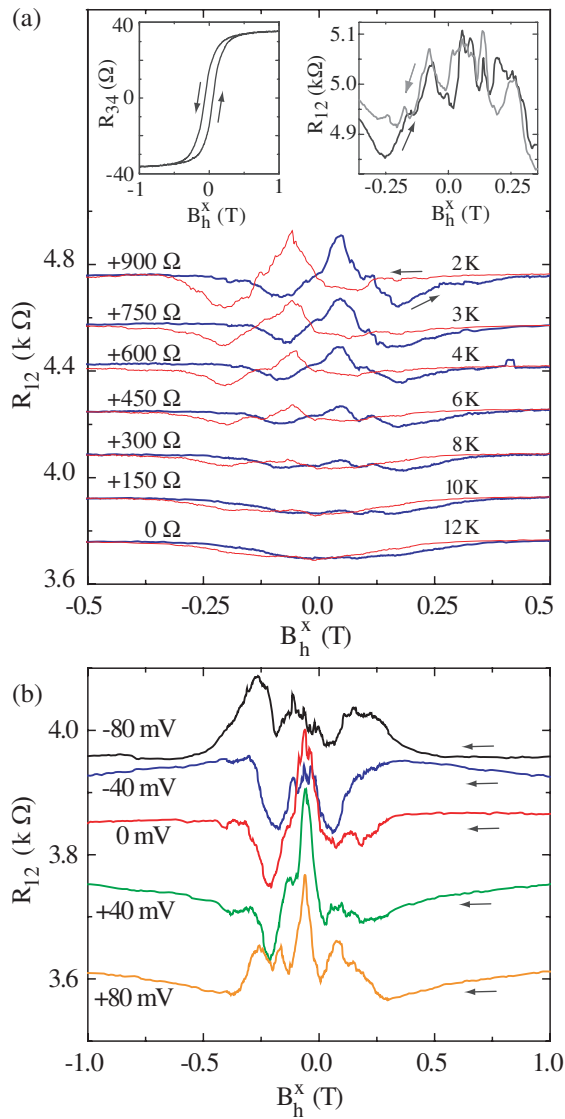}
\caption{(Color online) (a) Resistance of the QWR with the MB as a function of  $B_{h}^{x}$ at various temperatures. The sweep direction of $B_{h}^{x}$ is indicated by the arrows. Adjacent traces belonging to different temperatures have been offset with respect to the traces at $T=12\,\mathrm{K}$ for clarity as indicated. Left inset: Hall resistance $R_{34}$ as a function of  increasing and decreasing $B_{h}^{x}$. Right inset: up-sweep of $B_{h}^{x}$ in a different cooldown,
in comparison with the corresponding down-sweep, reflected about the $B_{h}^{x}=0$ axis. (b) Influence of  the in-plane gate voltages on the magnetoresistance of the QWR. }\label{MBCFinQWR_fig2}
\end{figure}

In Fig. \ref{MBCFinQWR_fig2}, the resistance of the QWR as a function of
increasing  $B_{h}^{x}$ is shown for various
temperatures.
At $12\,\rm{K}$, the magnetoresistance resembles that one of a MB in a diffusive 2DEG \cite{Vancura2000,Cerchez2007} with a
minimum at the coercive magnetic field of the Co film, which is determined by the vanishing of the Hall resistance $R_{34}$, see the left inset in Fig. \ref{MBCFinQWR_fig2}. \cite{Cerchez2007} As
the temperature is reduced, pronounced magnetoresistance
fluctuations appear. They are reproducible under sweeps of $B_{h}^{x}$ in the same direction, but the fluctuation pattern is modified under thermal cycling to room temperature (right inset). Thermal cycling can also change the QWR resistance by as much as 30\%, indicating a high sensitivity to the specific configuration of the scatterers.  In many, but not in all cooldowns, $R_{12}$ shows a maximum of varying amplitude and shape at the coercive magnetic field, which resembles a weak localization peak. The right inset of Fig. \ref{MBCFinQWR_fig2} (a) furthermore shows the magnetoresistance observed in an up-sweep at $2.1\,\rm{K}$  in comparison to the corresponding down-sweep, reflected about $B_{h}^{x}=0$.  Most features look very similar in both traces, indicating that they are invariant under inversion of the MB, as expected from symmetry arguments. \cite{Datta1997} Possible reasons for  the difference between these two traces are discussed at the end of this Section. The fluctuation pattern can be also tuned by the gate voltages. In Fig. \ref{MBCFinQWR_fig2} (b), $R_{12}(B_{h}^{x})$ is reproduced for various voltages applied to the in-plane gates $V_{\mathrm{IPG}}$. As $V_{\mathrm{IPG}}$ is reduced, the overall resistance increases due to the reduction of the electron density, while the fluctuation pattern changes non-monotonously. Our tuning range, however, is limited due to leakage currents across the oxide lines for $|V_{\mathrm{IPG}}|>80\,\mathrm{mV}$.

\begin{figure}
\includegraphics{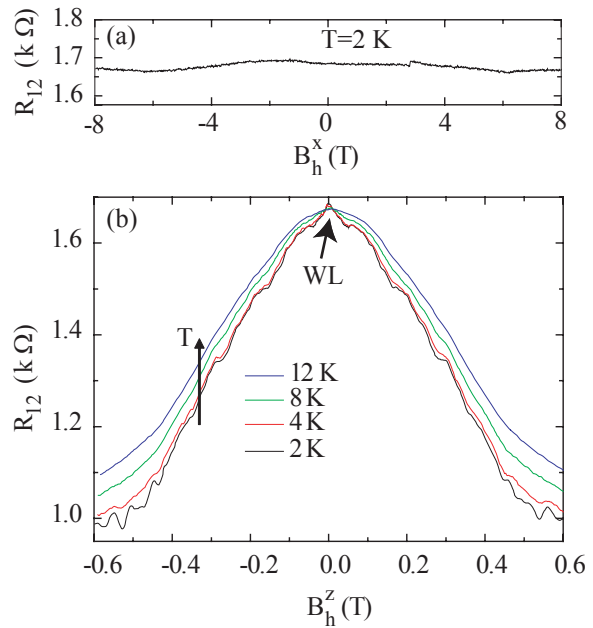}
\caption{(Color online) resistance of a QWR as shown in Fig. \ref{MBCFinQWR_fig1} (b) in homogeneous magnetic fields as a function of a homogeneous parallel (a) and perpendicular (b) magnetic field. The weak localization (WL) peak in (b) is denoted by the arrow.} \label{MBCFinQWR_fig3}
\end{figure}
We emphasize that our observations differ distinctly from those measured on QWRs in homogeneous magnetic fields. \cite{Nakamura1992,Nikolic1994,Held1999} For control purposes, we also measured the resistance of a QWR without a magnetic film as a function of both $B_{h}^{x}$ and $B_{h}^{z}$, see Fig. \ref{MBCFinQWR_fig3}. Even though this QWR is nominally identical to that one shown in Fig. \ref{MBCFinQWR_fig1}, its resistance is about a factor of 2.5 smaller. We attribute this  to the well-known fact that the lateral depletion length of the oxide lines depends sensitively on the oxidation depth \cite{May2007} leading to poor reproducibility. In the parallel configuration, the resistance is free of fluctuations and approximately independent of  $B_{h}^{x}$ (Fig. \ref{MBCFinQWR_fig3}(a)),
while the magnetoresistance in the perpendicular configuration shows the well-known behavior. \cite{Beenakker1991,Nakamura1992,Nikolic1994} The most prominent feature is a negative magnetoresistance with a weak temperature dependence. A superimposed weak localization peak at zero magnetic field is seen. In addition, magnetoresistance fluctuations with an amplitude of $\approx 10\,\mathrm{\Omega}$ at $2.0\,\mathrm{K}$ corresponding to a conductance fluctuation amplitude of $\delta G\approx 0.08 e^2/h$, are visible. We will comment on the different magnetoresistance features in homogeneous vs. localized magnetic fields below, subsequent to the discussion of the numerical simulations.

Furthermore, our system should also be distinguished from the wires investigated by Hara et al. \cite{Hara2004a}, where resistance fluctuations as a function of an inhomogeneous magnetic field in a wire were observed as well. The magnetic field pattern in this experiment consists of a strong gradient in y-direction but is constant in longitudinal direction, whereas the electrons in our QWRs see a localized magnetic field in transport direction but homogeneous in y-direction.

\begin{figure}
\includegraphics{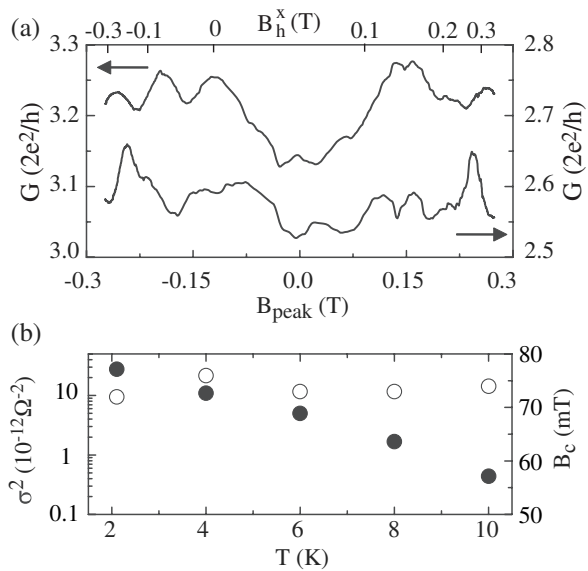}
\caption{(a) Conductance G of the QWR as a function of the peak
magnetic field of the barrier, as obtained from two cooldowns. The (nonlinear) axis on top specifies the corresponding values of $B_h^x$. (b) $\sigma^2$ (full circles) and correlation magnetic field (empty circles) of $G$
of the data in Fig. \ref{MBCFinQWR_fig2}(a) as a function of  temperature.} \label{MBCFinQWR_fig4}
\end{figure}

For a more quantitative characterization of the MB induced resistance fluctuations, we map $B_h^x$ onto $B_{peak}$ as a characteristic quantity. This is achieved by determining the magnetization of the Co film as a function of the applied magnetic field  $\mu_0 M(B_{h}^{x})$ via Hall magnetometry. As described in detail in Refs.  \cite{Vancura2000, Cerchez2007}, the measured Hall resistance $R_{34}(B_{h}^{x})$, shown in the left inset in Fig. \ref{MBCFinQWR_fig2}(a), allows to determine $\mu_0 M$ which leads to $B_{peak}$ via Eq. \eqref{eqn1}. This procedure assumes identical magnetization
characteristics at both edges, which has been shown to be the case to high accuracy in earlier experiments. \cite{Cerchez2007} The conductance $G\equiv R_{12}^{-1}$ as a
function of $B_{peak}$ is plotted in Fig. \ref{MBCFinQWR_fig4} (a) for the two cooldowns at $2\,\mathrm{K}$ shown in Fig \ref{MBCFinQWR_fig2} (a). A broad minimum around $B_{peak}=0$
of variable markedness is observed, while the conductance fluctuations extend over the whole range of  $B_{peak}$. In Fig. \ref{MBCFinQWR_fig4} (b), we show the temperature dependence of both the amplitude $\sigma^2=\mathrm{var} (G)$  after subtraction of a smooth background, and the correlation magnetic field  $B_{c}$. The corresponding amplitude of the conductance fluctuations at $2\,\mathrm{K}$ equals $\delta G=\sigma=0.15\,e^2/h$, which is a factor of 2 above that value for homogeneous magnetic fields. While $B_c\approx 75\,\mathrm{ mT}$ is temperature independent below $12\,\mathrm{K}$,  $\sigma^2$ decays
approximately exponentially with increasing temperature, in the temperature range of our experiment. Comparable quantum wires in homogeneous magnetic fields show also a temperature independent $B_c$, but $\sigma^2$ decays algebraically. \cite{Nakamura1992}  Further experiments, in particular at lower temperatures, as well as a detailed theoretical study of MBs in quasi-ballistic quantum wires are probably required for a better understanding of this behavior.

We proceed by developing a qualitative interpretation of our observations and support it by a numerical model based on the recursive Green's functions technique. The reproducibility of the fluctuations and their strong temperature dependence suggest a quantum origin. We therefore
interpret them as a coherence effect tuned by the MB. As the electrons get scattered at the potential landscape formed by impurities and the wire edges, the coherent part of the electron wave function generates an interference pattern which depends sensitively not only on the configuration of the scatterers,
but also on the magnetic field. \cite{Beenakker1991} The resulting magnetoconductance
patterns are also known as magneto-fingerprints of the sample and are
usually not universal in quantum wires. \cite{Nikolic1994} In our
system, the magnetic phase collected by the electron waves depends strongly on $x$. As $B_{peak}$ is varied, the magnetic phase shift is most significant in those random resonators located in close proximity to the MB. Since only a few such resonators exist, the shape and strength of the weak localization peak depends on the configuration of the scattering centers.

\begin{figure}
\includegraphics{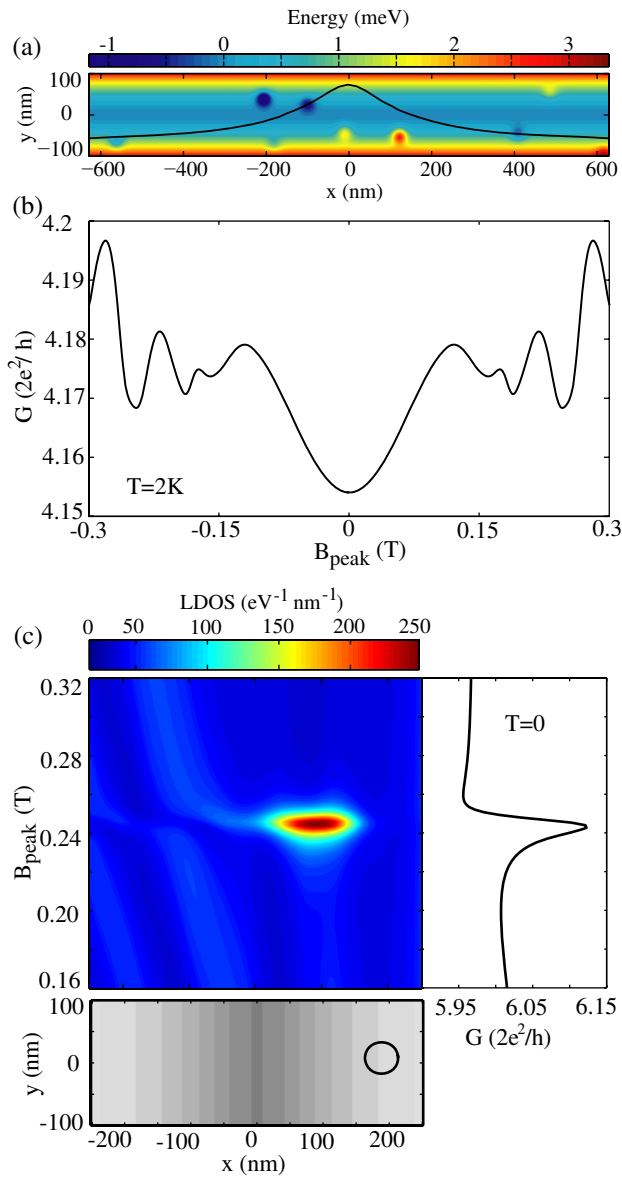}
\caption{(Color online) A typical disorder potential used in the simulations (a). The MB is indicated by the black line. (b) The corresponding conductance as a function of $B_{peak}$ at a temperature of $2\,\mathrm{K}$. (c) Left: the local density of states (LDOS) along the QWR containing one impurity (its position relative to the MB is shown in the lower part, where its radius at half maximum is indicated by the circle, and  the barrier shape by the gray tone) as a function of the MB strength. The LDOS is integrated over the width of the QWR. Right: the corresponding conductance of the QWR. }
\label{MBCFinQWR_fig5}
\end{figure}

In order to substantiate this picture, we calculate the conductance of a corresponding model system as a function of the MB strength such that it can be compared to the data shown in Fig. \ref{MBCFinQWR_fig3} (a). The QWR is modeled by a parabolic confinement potential $V(y)=\frac{1}{2}m^*\omega_0^2 y^2$ with $\hbar\omega_0=1.58\,\mathrm{meV}$ and a length of $L =4\,\mathrm{\mu m}$. The Fermi energy was set to $11\,\mathrm{meV}$, and a MB of the shape given by Eq. \eqref{eqn1} with $h_0=250\,\mathrm{nm}$ and $z_0=65\,\mathrm{nm}$ was used. These values are consistent with the information about the QWR that could be extracted from the experiment. Elastic scatterers are modeled by circular symmetric potentials of a Gaussian shape and a full width at half maximum of $30\,\mathrm{nm}$. The amplitudes $eV_0$ of the scatterers follow a Gaussian distribution centered around $eV_0=0$ with a half width at half maximum of $5\,\mathrm{meV}$. These scatterers are distributed in the QWR at random positions with a reasonable density of $0.33\,\mathrm{\mu m}^{-2}$, corresponding to an average separation between scatterers of $1.7\,\mathrm{\mu m}$. The resulting potential landscape of one scatterer configuration, depicted in Fig. \ref{MBCFinQWR_fig5} (a), is similar to those obtained within self-consistent models for comparable QWRs, see, e.g. Figs. 2 and 3 in Ref. \cite{Evaldsson2008}.

 The system is described by the Schr\"{o}dinger equation
\begin{equation}
\left[ H_0 + \frac{1}{2}m^*\omega_0^2 y^2 +V^{imp} \right]\psi(x,y) = E\psi(x,y)
\end{equation}
where $H_0$ is the kinetic energy term,  and
$V^{imp}$ is the potential due to impurities. Choosing the Landau gauge,
$\mathbf{A}=(-B_z(x)y,0,0)$, the kinetic energy part can be further written as
\begin{equation}
H_0=-\frac{\hbar^2}{2m^*}\left\{ \left( \frac{\partial}{\partial
x}-\frac{ieB_z(x)y}{\hbar} \right)^2 + \frac{\partial ^2}{\partial y^2}\right\}
\end{equation}

In order to perform numerical computations, the QWR area is
discretized into a grid lattice with lattice constant $a=3\,\mathrm{nm}$ such that the
continuous quantities $x$ and $y$ are replaced by discrete variables $ma$ and
$na$, respectively. Using the Peierls substitution, the magnetic field is included via a
phase factor in the hopping amplitudes, we arrive at the
tight-binding Hamiltonian
\begin{widetext}
\begin{eqnarray}
  H = \sum_{m,n}\bigg\{|m,n\rangle\left(\epsilon_0+ \frac{1}{2}m^*\omega_0^2a^2n^2  +V_{mn}^{imp}\right) \\ \langle m,n| -t\bigg(|m,n\rangle \langle m,n+1|\nonumber  &+&|m,n\rangle e^{-iqna}\langle m+1,n| + H.c.\bigg) \bigg \}
\end{eqnarray}
\end{widetext}
where $t=\hbar^2/(2m^*a^2)$ is the nearest-neighbor hopping element,
$\epsilon_0=4t$ is the site energy, and $q=\frac{e}{\hbar} \int^{x_{i+1}}_{x_i}
B_z(x')dx'$. In the calculation, the standard recursive technique is used to
compute the total Green's function which is related to the transmission
amplitude from mode $\alpha$ to $\beta$ via the expression
$t_{\beta\alpha}=i\hbar\sqrt{v_{\alpha} v_{\beta}}\mathbf{G}^{M+1,0}$, where
$v_{\alpha(\beta)}$ is the group velocity, and $\mathbf{G}^{M+1,0}$ denotes the
matrix $\langle M+1|\mathbf{G}|0 \rangle$, with 0 and $M+1$ corresponding to
the positions of the left and right leads. We calculate separately the surface
Green's functions related to the left and right leads and link them to the
Green's function of the scattering region with a MB.

The two-terminal conductance $G(E)$ is calculated within the
framework of Landauer-B\"uttiker formalism
\begin{equation}
    G(E)=\frac{2e^2}{h}\sum_{\alpha,\beta=1}^N|t_{\beta\alpha}|^2(E)
\end{equation}
with $N$ being the number of propagating states in the leads, from which we finally obtain the conductance at temperature $T$ according to $G=\int G(E)(-\frac{df(E,T)}{dE}) dE$, where $f(E,T)$ denotes the Fermi-Dirac distribution function.

In Fig. \ref{MBCFinQWR_fig5} (b), we show the calculated magnetoconductance $G(B_{peak})$ at a temperature of $2\,\mathrm{K}$. Both the conductance fluctuations and the broad minimum around $B_{peak}=0$ seen in the measurements are qualitatively reproduced, indicating that a varying MB can indeed be the origin of the observed phenomenology. Quantitative differences remain. In particular, the simulated conductance is a factor of  $\approx 1.4 $ above the measured ones (Fig. \ref{MBCFinQWR_fig3} (a)). Also, the fluctuation amplitude is only $\approx 0.02 e^2/h$, much smaller than the experimental value, while the width of the conductance minimum around $B_{peak}=0$ corresponds roughly to the observed one. A more quantitative agreement would require a self-consistent simulation of the QWR potential landscape. \cite{Evaldsson2008} This however is beyond the scope of the present paper, and we note that due to the high sensitivity of the wire parameters on the details of the scanning probe lithography, \cite{May2007} as well as the strong changes under thermal cycling, a full quantitative agreement may be difficult to achieve.

Furthermore, the developed model can be used to shed some light on the character of the conductance fluctuations, which are expected to be absent in a clean, ballistic QWR with a MB. \cite{Xu2007a} The origin of the fluctuations is exemplified by a parabolic QWR containing a tunable MB and  a single repulsive scatterer close-by, of the above described shape and of an amplitude $eV_0=5\,\mathrm{meV}$, see Fig. \ref{MBCFinQWR_fig5} (c). At $B_{peak}=250\,\mathrm{mT}$, the conductance shows an asymmetric resonance with a Fano character. \cite{Vargiamidis2005} The corresponding local density of states (LDOS) shows a peak at this value of $B_{peak}$, localized in between the scatterer and the center of the MB. This indicates that the MB
acts as a repulsive scatterer which forms one mirror of a resonator for electron waves. Further simulations (not shown) reveal that the position and the shape of such resonances
depend sensitively on the position of the scatterer, and both peaks as well as dips in $G(B_{peak})$ are observed as the position of the scatterer is varied. Moreover, the character of the resonances does not change as the sign of the scattering potential is reversed. We note that this type of resonance is the only one we could identify in our simulations, suggesting that the magnetoresistance fluctuations originate from a superposition of such resonances.

With this interpretation in mind, it is insightful to return to the comparison of the magnetoconductance of quantum wires in homogeneous vs. localized perpendicular magnetic fields (Figs. \ref{MBCFinQWR_fig2} and \ref{MBCFinQWR_fig3} (b)). First of all, the negative magnetoresistance in homogeneous magnetic fields is absent in QWRs with a MB in its center. This is easily understandable since this effect originates from a magnetic field induced reduction of electron reflections at the entrance of the QWR. \cite{Houten1992} At these points, however, the magnetic field of the MB is negligible. Second, the weak localization peak observed in homogeneous magnetic fields \cite{Beenakker1991} is not always observed in QWRs with MBs and has no characteristic shape. We speculate that the field of the MB acts as an x-dependent phase shifter for states which are weakly localized by scattering at the impurities. How exactly weak localization in
QWRs exposed to localized magnetic fields modifies the conductance remains to be studied in future theoretical work. However, a heuristic argument delivers a plausible explanation for the width of the broad conductance dip. In homogeneous magnetic fields, the half width at half
maximum $B_{1/2}$ of  the weak localization dip corresponds to a characteristic area $A= \hbar/(2eB_{1/2})$. \cite{Beenakker1991,Ouchterlony1999,Zozoulenko1996c}  We observe a conductance dip of width $B_{peak,1/2}\approx 80\,\mathrm{mT}$, corresponding to an average magnetic field in the QWR of $9.5\,\mathrm{mT}$.  The characteristic area is thus $A= 3.5\times 10^{4}\,\mathrm{nm}^2$. Assuming a wire width of $\approx 200\,\mathrm{nm}$, a characteristic length of $175\,\mathrm{nm}$ is obtained, which appears realistic for the average extension of a backscattering loop along the QWR. Furthermore, the resistance fluctuation amplitude at identical sample mobilities and temperatures is enhanced in the samples with the MB. Qualitatively, this can be understood along the same lines as cooldown-dependent, irregularly shaped  weak localization dip: since the section of the QWR which is tuned by the MB is much shorter than its length, the averaging of the conductance fluctuations is reduced, leading to larger fluctuation amplitudes.

\begin{figure}
\includegraphics{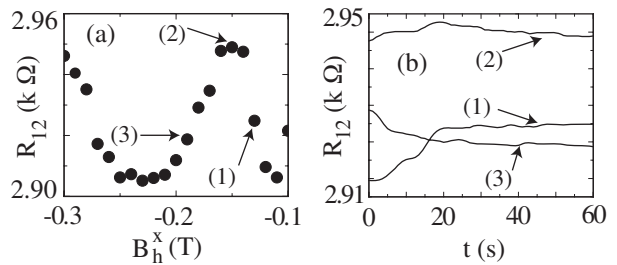}
\caption{Section of the QWR resistance as a function of $B_h^x$ after $60\,\mathrm{s}$ holding time in each point (a) and the time dependence of $R_{12}$ at points 1 to 3  (b). }
\label{MBCFinQWR_fig6}
\end{figure}

Finally, we would like to dwell on the measured deviations from the expected symmetry relation $R_{12}(B_{peak})=R_{12}(-B_{peak})$. In the right inset in Fig. \ref{MBCFinQWR_fig2} (a), one observes resistance differences up to $100\,\mathrm{\Omega}$ for some magnetic fields, while the main features are present in both up and down sweeps. In order to detect a possible time dependence of $R_{12}$, we have
changed $B_h^x$ from $+2\,\mathrm{T}$ to $-2\,\mathrm{T}$ in $10\,\mathrm{mT}$ steps, and measured
$R_{12}(B_h^x,t)$ up to a time of $t=60\,\mathrm{s}$ for each step. Fig. \ref{MBCFinQWR_fig6}(a) shows  $R_{12}(B_h^x)$ for  $t=60\,\mathrm{s}$. In Fig. \ref{MBCFinQWR_fig6}(b), the evolution of $R_{12}(B_h^x)$ over time at three values of $B_h^x$ are reproduced.  One point each was chosen to the right (1) and to the left (3) of a  local resistance maximum, where the susceptibility to
small variations of  the magnetization is high, and one point (2) near a local resistance maximum. The resistance at point (1) increases by $\approx 14\,\mathrm{\Omega}$ in  $20\,\mathrm{s}$, while at point (2),  the change in magnetization drives the wire resistance through
the local maximum with a variation over time of $5\,\mathrm{\Omega}$ only, and in point (3), the resistance drops by  $\approx 9\,\mathrm{\Omega}$ over $20\,\mathrm{s}$. Since the time constant of the low-pass filter in our measurement setup is set to $1\,\mathrm{s}$, these observations cannot be explained by external effects. Rather, we attribute them to changes in the film magnetization with time due to thermal activation over local
energy barriers, also known as \emph{magnetic aftereffect}, which are reported to show a similar time dependence in other Co films. \cite{Ferre1997} Unfortunately, these time-dependent changes in $R_{12}$
cannot be correlated to those observed in the hysteresis loop ($R_{34}$), since the QWR probes the edge of the Co film locally, while the Hall sensing averages over the edge on its opposite side. Hence, even though magnetic relaxation effects do contribute to the asymmetry of $R_{12}$, their amplitude in resistance is significantly  smaller than the maximum deviations observed between up- and down-sweeps of $B_h^x$. Therefore, we believe that the asymmetry originates from both background charge rearrangements in the semiconductor as well as from magnetic relaxation in the ferromagnet.

\section{\label{sec:4}SUMMARY AND CONCLUSIONS}

In summary, we have presented  experimental and numerical studies regarding the transport properties of a quasi-ballistic quantum wire exposed to a highly localized perpendicular magnetic field. The magnetoresistance of this system differs distinctly from that one known from quantum wires exposed to homogeneous magnetic fields: the negative magnetoresistance is absent, while the amplitude of the conductance fluctuations is enhanced. In addition, a broad minimum in the magnetoconductance is observed and interpreted as an indication of weak localization. Within a recursive Green's function model, it is found that the conductance fluctuations originate from electronic interferences between electrostatic scatterers and the magnetic barrier.  We hope that these findings will motivate further theoretical studies to elucidate the physics of this system in quantitative terms.\\

M. C. and H. X. acknowledge financial support from the Forschungs-F\"orderungsfonds of the HHU D\"usseldorf. H. X., I. Z. and T. H. acknowledge support by the DAAD via the DAAD-STINT programme; A. D. W. and D. R. acknowledge financial support within the BMBF nanoQUIT and the SFB 491.


\end{document}